\documentclass[sigconf]{acmart}
\acmConference[Conference 2024]{ACM Conference}{X, 2024}{X, X}
\pdfoutput=1
\setcopyright{acmcopyright}
\copyrightyear{2018}
\acmYear{2018}
\acmDOI{XXXXXXX.XXXXXXX}

\acmPrice{15.00}
\acmISBN{978-1-4503-XXXX-X/18/06}

\renewcommand\footnotetextcopyrightpermission[1]{}

\usepackage{comment}
\usepackage{pifont}
\usepackage{pbox}
 \usepackage[flushleft]{threeparttable}

\usepackage{booktabs}

\usepackage{diagbox}
\usepackage{tablefootnote}

\usepackage{subcaption}

\usepackage{hyperref}
\usepackage{algorithmic}
\usepackage{adjustbox}
\usepackage{multirow}
\usepackage{tabularx}
\usepackage{enumitem}
\usepackage{framed}
\usepackage[linesnumbered,ruled]{algorithm2e}
\SetKwInput{KwInput}{Input} \SetKwInput{KwOutput}{Output} \SetKwInput{KwProcedure}{Procedure}
 \SetCommentSty{mycommfont}
\usepackage[colorinlistoftodos]{todonotes}
\usepackage{graphicx}
\usepackage{textcomp}
\usepackage{xcolor}
\usepackage{multicol}
\usepackage{balance}
\usepackage{subcaption}
\usepackage{epstopdf}
 \AppendGraphicsExtensions{.pdf}
\usepackage{array}
\usepackage{listings}
\definecolor{Gray}{gray}{0.9}
\definecolor{pgreen}{rgb}{0,0.5,0}
\usepackage{amsthm}
\makeatletter
\def\th@plain{%
  \thm@notefont{}
  \itshape 
}
\def\th@definition{%
  \thm@notefont{}
  \normalfont 
} \makeatother
\usepackage{capt-of}

\newtheorem{definition}{Definition}

\usepackage{color, colortbl}
\definecolor{grey}{rgb}{0.7,0.7,0.7}

\newcommand{\lstbg}[3][0pt]{{\fboxsep#1\colorbox{#2}{\strut #3}}}
\lstdefinelanguage{diff}{
  basicstyle=\ttfamily\scriptsize,,
  morecomment=[f][\lstbg{red!20}]-,
  morecomment=[f][\lstbg{green!20}]+,
  morecomment=[f][\lstbg{yellow!20}]++,
  morecomment=[f][\textit]{@@},
  texcl=false
}

\usepackage{tcolorbox}
\usepackage{xcolor}
\definecolor{todocolor}{rgb}{0.9,0.1,0.1}
\definecolor{indiagreen}{rgb}{0.07, 0.53, 0.03}
\definecolor{hycolor}{rgb}{0.7,0.7,0.3}
\definecolor{darkbrown}{rgb}{0.4, 0.26, 0.13}

\usepackage{listings}
\usepackage{xcolor}
\definecolor{main-color}{rgb}{0.6627, 0.7176, 0.7764}
\definecolor{string-color}{rgb}{0.3333, 0.5254, 0.345}
\definecolor{key-color}{rgb}{0.8, 0.47, 0.196}

\lstdefinestyle{mystyle} {
    language = Java,
    basicstyle = {\ttfamily \color{main-color}},
    stringstyle = {\color{string-color}},
    keywordstyle = {\color{key-color}},
    keywordstyle = [2]{\color{lime}},
    keywordstyle = [3]{\color{yellow}},
    keywordstyle = [4]{\color{teal}},
    morekeywords = [3]{<<, >>},
    morekeywords = [4]{++},
    basicstyle=\ttfamily\scriptsize,
    commentstyle=\color{blue}\ttfamily,
    morecomment=[f][\lstbg{red!20}]-,
    morecomment=[f][\lstbg{green!20}]+,
    morecomment=[f][\lstbg{yellow!20}]++,
    morecomment=[f][\lstbg{yellow!20}]--,
    morecomment=[f][\textit]{@@},
    texcl=false
}

\lstnewenvironment{bash}[1][]
  {\lstset{language=[latex]TeX}\lstset{%
   numbers=left,numberstyle=\scriptsize,
   commentstyle=\color{pgreen},
   xleftmargin=5pt,
   xrightmargin=5pt,
   aboveskip=\bigskipamount,
   belowskip=\bigskipamount, #1,
}} {}

\lstdefinestyle{testlstcolor}{
    language={sh},
    moredelim=**[is][\color{red}]{~}{~},
    moredelim=**[is][\color{blue}]{<}{>},
    moredelim=**[is][\bfseries]{***}{***},
    moredelim=**[is][\color{green}]{~~}{~~},
    showstringspaces=false,
    basicstyle=\ttfamily,
    literate={\\~}{{\textasciitilde}}1
        {\\<}{{\unichar{"003C}}}1
        {\\>}{{\unichar{"003E}}}1
}

\newcolumntype{L}[1]{>{\raggedright\let\newline\\\arraybackslash\hspace{0pt}}m{#1}}
\usepackage{tikz}

\usepackage{xspace}

\newcommand\cparagraph[1]{\noindent\textbf{#1.}\xspace}

\newcommand{\tooln}{\textsc{SimpT5}\xspace}
\newcommand{\bench}{\textsc{SimpliBench}\xspace}
\newcommand{\jdeo}{\textsc{JDeodorant}\xspace}

\newcommand{\tufano}{\textsc{TufanoNMT}\xspace}
\newcommand{\autotransform}{\textsc{AutoTransform}\xspace}

\newcommand{\totnumpairbench}{92,485\xspace} 

\newcommand{\numbench}{329\xspace} 
\newcommand{\numprojbench}{307\xspace} 

\newcommand{\exactmatch}{30\xspace} 

\newcommand{\studycohenkappa}{0.81\xspace} 

\newcommand{\Plus}{\mathord{\begin{tikzpicture}[baseline=0ex, line width=0.5, scale=0.06]\draw (1,0) -- (1,2);
\draw (0,1) -- (2,1);
\end{tikzpicture}}}
\newcommand{\Minus}{\mathord{\begin{tikzpicture}[baseline=0ex, line width=0.5, scale=0.06]
\draw (0,1) -- (2,1);
\end{tikzpicture}}}
\def\HiLir{\leavevmode\rlap{\hbox to \hsize{\color{red!50}\leaders\hrule height .8\baselineskip depth .5ex\hfill}}}
\def\HiLi{\leavevmode\rlap{\hbox to \hsize{\color{blue!50}\leaders\hrule height .8\baselineskip depth .5ex\hfill}}}

\newcolumntype{H}{>{\setbox0=\hbox\bgroup}c<{\egroup}@{}}

\newcolumntype{x}[1]{%
{\centering\hspace{0pt}}p{#1}}%

\newcommand{\delcatstudy}{16.49}

\newcommand{\simpli}{developer-induced program simplification}
\newcommand{\csimpli}{Developer-induced Program Simplification}
\newcommand{\cpsimpli}{Developer-induced program simplification}
\definecolor{azure(colorwheel)}{rgb}{0.0, 0.5, 1.0}
\newcommand{\modifyhaibo}[1]{{\color{azure(colorwheel)}{#1}}}

\newcommand{\useapip}{Replace with equivalent API}
\newcommand{\useapi}{``\useapip''\xspace}

\newcommand{\refactoringminer}{RefactoringMiner}

\newcommand{\analyzedPRs}{382}
\newcommand{\analyzedRepos}{296}
\newcommand{\analyzedMotivationPRs}{79}
\newcommand{\numberofmaincategories}{seven}
\newcommand{\numberofsubcategories}{26}
\newcommand{\numberofmotivations}{four}

\newcommand{\readabilityratio}{26.58}

\newcommand{\complexityratio}{8.86}

\newcommand{\deletionratio}{62.03}

\newcommand{\reusabilityratio}{3.80}

\newcommand{\controllogicrelatedratio}{39.53}

\newcommand{\functionreturnratio}{13.87}
\newcommand{\conditionalexpressionnumber}{32}
\newcommand{\conditionalexpressionratio}{8.42}

\newcommand{\extractionratio}{31.41}

\newcommand{\extractmethodratio}{19.63}

\newcommand{\apiratio}{16.23}

\newcommand{\diamondnumber}{16}
\newcommand{\diamondratio}{4.19}

\newcommand{\trywithnumber}{2}
\newcommand{\trywithratio}{0.52}

\newcommand{\languagefeatureratio}{13.09}

\newcommand{\bigdatasetreponumber}{25022}
\newcommand{\bigdatasetcommitnumber}{37846}

\newcommand{\validaveragetestnumber}{96}

\newcommand{\unchangeditemratio}{30}
\begin{document}

\title{Moving beyond Deletions: Program Simplification via Diverse Program Transformations}




\author{Haibo Wang}
\affiliation{%
  \institution{Concordia University}
  \city{Montreal}
  \country{Canada}}
\email{haibo.wang@mail.concordia.ca}

\author{Zezhong Xing}
\affiliation{%
  \institution{Southern University of Science and Technology}
  \city{Shenzhen}
  \country{China}}
\email{12232384@mail.sustech.edu.cn}

\author{Zheng Wang}
\affiliation{%
  \institution{University of Leeds}
  \city{Leeds}
  \country{UK}}
\email{z.wang5@leeds.ac.uk}

\author{Chengnian Sun}
\affiliation{%
  \institution{University of Waterloo}
  \city{Waterloo}
  \country{Canada}}
\email{cnsun@uwaterloo.ca}

\author{Shin Hwei Tan}
\affiliation{%
  \institution{Concordia University}
  \city{Montreal}
  \country{Canada}}
\email{shinhwei.tan@concordia.ca}

\begin{abstract}
To reduce the complexity of software, Developers manually simplify program (known as \simpli{} in this paper) to reduce its code size yet preserving its functionality but manual simplification is time-consuming and error-prone.
To reduce manual effort, rule-based approaches (e.g., refactoring) and deletion-based approaches (e.g., delta debugging) can be potentially applied to automate \simpli{}.
However, as there is little study on how developers simplify programs in Open-source Software (OSS) projects, it is unclear whether these approaches can be effectively used for \simpli{}. 
Hence, we present the first study of \simpli{} in OSS projects, focusing on the types of program transformations used, the motivations behind simplifications, and the set of program transformations covered by existing refactoring types. Our study of \analyzedPRs{} pull requests from \analyzedRepos{} projects reveals that there exist gaps in applying existing approaches for automating \simpli{}. 
and outlines the criteria for designing automatic program simplification techniques. 
Inspired by our study and to reduce the manual effort in \simpli{},
we propose \tooln{}, a tool that can automatically produce simplified programs (semantically-equivalent programs with reduced source lines of code). \tooln{} is trained based on our collected dataset of \totnumpairbench{} simplified programs with two heuristics: (1) \emph{simplified line localization} that encodes lines changed in simplified programs, and (2) checkers that measure the quality of generated programs. Our evaluation shows that \tooln{} are more effective than prior approaches in automating \simpli{}. 
\end{abstract}


\ccsdesc[500]{Software and its engineering~General programming languages}

\keywords{Program Simplification, Program Reduction, Refactoring}

\maketitle
\section{Introduction}
\label{sec:intro}

Software systems have become increasingly more complex. To ease software maintenance, developers often dedicate significant time to simplify programs manually -  a process referred to as \emph{\simpli} in this paper. The goal is to reduce the code size or the number of lines of code of the programs while preserving their functionalities. 
 As defects can be introduced when modifying programs by hand~\cite{ge2012reconciling}, manual program simplification is error-prone. Meanwhile, several techniques can potentially be applied for automating \simpli{}: (1) rule-based approaches such as refactoring that restructures code via a predefined set of \emph{syntactic program transformations} ~\cite{becker1999refactoring} (also known as \emph{refactoring types}), and (2) deletion-based techniques such as delta
debugging~\cite{ddinput} that performs \emph{semantic program simplification} by relying on test executions to check whether removing parts of the input program leads to a smaller program that is semantically equivalent to the original program. 

Refactoring, a representative of rule-based approaches,  has been extensively studied. Most prior research efforts in this direction focus either on investigating the refactoring practice ~\cite{oliveira2023untold,ge2012reconciling,kim2012field,paixao2020behind,peruma2022refactor,alomar2019can} or designing automated refactoring techniques~\cite{gyori2013crossing,binkley2005automated,meng2015does,jensen2010use,karakati2022software}. However, the findings of these studies and the effectiveness of prior refactoring techniques may not be applicable in the context of \simpli{} because (1) they mainly focus on a limited set of supported refactoring types (e.g., RefactoringMiner can detect 99 refactoring types), and (2) some refactoring types may not have the desired properties of a simplified program (having reduced code size while preserving the program behavior). For example, ``Extract Local Variable'', a popular refactoring type, may introduce an additional line of code due to the insertion of a new variable.  


Meanwhile, deletion-based techniques have demonstrated promising results in automatically producing reduced programs that can be used as inputs for several domains, including debugging~\cite{ddmin,hammoudi2015use}, test case simplification~\cite{zhang2013practical,ddinput,jiang2017simplydroid}, and improving the understanding of features in neural code intelligence system~\cite{rabin2022syntax}.
Understanding the types of program transformations used in \simpli{} is essential as it can potentially enhance these domains where deletion-based techniques are used by producing more diverse reduced programs. 

\begin{table*}
\centering
\caption{Differences with relevant work that aim to produce reduced programs}
\label{tab:Differences}
\small
\begin{adjustbox}{width=0.98\textwidth}
\begin{tabular}{l|l|l|l|l} 
\hline
\multicolumn{1}{c|}{Approach} & \multicolumn{1}{c|}{Simplification Goals}             & \multicolumn{1}{c|}{Transformations}                  & \multicolumn{1}{c|}{Simplification Technique}                                           & \multicolumn{1}{c}{Tests}  \\ 
\hline\hline
Delta Debugging$^D$           & Debugging, test input minimization, feature isolation & Deletion                                              & \begin{tabular}[c]{@{}l@{}}Delta debugging~\cite{ ddmin,ddinput},\\hierarchical delta debugging~\cite{misherghi2006hdd, hodovan2017coarse}\end{tabular} & Need to run tests          \\ 
\hline
Program Reduction$^D$         & Test case minimization                                & Deleting/Replacing bug-irrelevant elements                      & Perses~\cite{sun2018perses}, Vulcan~\cite{xu2023pushing}                                                                         & Need to run tests          \\ 
\hline
Program Debloating$^D$        & Reduce attack surfaces                                & Deletion                                              & Combine with delta debugging~\cite{heo2018effective}, Dependency Analysis~\cite{qian2019razor}                                      & Need to run tests          \\\hline 
Program Slicing$^D$           & Debugging, testing, software maintenance              & Deletion* & Dependency Analysis~\cite{agrawal1990dynamic,harman1995using,lu2012automatic,binkley2013observation,wong2023slicing}                                                                    & May/May not run tests      \\ 
\hline
Simplication in GP$^R$        & Enforce explanability of models                       & Rule-based Transformations                            & Simplification Rules (e.g., algebraic, numerical)
& May/May not run tests      \\ 
\hline
Refactoring$^R$               & Software maintenance                                  & Rule-based Transformations                            & Rules based on refactoring types~\cite{zhang2013practical,nievergelt1965automatic,ddinput,jiang2017simplydroid,tsantalis2020refactoringminer,Tsantalis:ICSE:2018:RefactoringMiner}                                                       & May/May not run tests      \\ 
\hline\hline
Our Work                      & Generate                                              & Diverse Transformations                               & Based on deep learning and LLMs                                                         & Need to run tests          \\
\hline
\end{tabular}
\end{adjustbox}
\begin{tablenotes}
\footnotesize
\item{
Approaches marked with $^D$ 
 represents deletion-based approaches, whereas approaches marked with $^R$ denotes rule-based approaches. *We consider program slicing as a deletion-based approach that removes statements without dependency from the original program.
}
\end{tablenotes}
\end{table*}


To fill in the gaps of prior studies and techniques, we present the first study of the characteristics of \simpli{} in open-source software (OSS) projects in GitHub. 
Specifically, our study 
investigated \emph{the types of program transformations used and the motivations behind \simpli{}} by answering the research questions below:
\begin{description}[leftmargin=*]
\item[RQ1] What are the program transformations commonly used in \simpli{}?
\item[RQ2] What are the motivations behind \simpli{} in OSS projects?
\item [RQ3] What are the program transformations covered by existing refactoring types?
\item[RQ4] How effective are prior tools for refactoring detection (RefactoringMiner) and refactoring automation (\jdeo) in detecting and automating \simpli?
\end{description}
As refactoring and \simpli{} aim to improve code while preserving program semantics, we investigated the feasibility of reusing existing tools for detecting and automating refactoring for \simpli{}.

Based on the findings of our study, we propose \tooln{}, a tool that automates \simpli{} by integrating (1) syntactic program transformations by using a diverse set of program transformations and (2) semantic program simplifications by using tests to check for test-equivalent relations. 

 In summary, we made the following contributions:
\begin{itemize}[leftmargin=*,labelindent=7pt,nosep]
    \item To the best of our knowledge, we present the first study of the program transformations used by developers and the motivations that drive developers when performing \simpli{}. The key findings of our study include (1) there are diverse types (\numberofsubcategories{} types) of program transformations used in \simpli{}, (2) several  frequently used simplification types have not been covered by deletion-based approaches and prior refactoring types (e.g., replacing a code block with an API call and simplifying expressions), (3) among the supported refactoring types, prior refactoring detection engine is still limited due to complex rule design, 
    (4) prior auto-refactoring tools fail to automate many simplifications, indicating the need for a new tool with a richer set of transformations. 
    \item We introduce \tooln, a proof-of-concept simplification framework based on pretrained large language model that uses two heuristics: (1) simplified line localization (encode line to be simplified), (2) checkers that evaluate the quality of generated programs via static metrics observed in our study.  
     \item We propose \bench, a benchmark containing \bigdatasetcommitnumber{} code simplification commits from \bigdatasetreponumber{} open-source Java projects. We also derive a \emph{valid} dataset from this benchmark, which contains \numbench{} simplification commits across
         \numprojbench{} projects. The \emph{valid} dataset contains commits that are compilable together with tests to validate the quality of the generated programs. Our evaluation on \bench{} shows that \tooln{} can automatically simplify code with \exactmatch{}\% exactly match rate. The dataset and the source code 
         are publicly available at ~\cite{dataandcodelink}.
\end{itemize}

\section{Related Work and Problem Formulation}
\label{sec:related}

Table~\ref{tab:Differences} highlights the differences between our work and prior techniques that produce programs of reduced sets.
The problem of \emph{automated program simplification} was widely studied~\cite{zhang2013practical,ddinput,jiang2017simplydroid} with the earliest work dated back to 1965~\cite{nievergelt1965automatic}.  We briefly discuss prior definitions of program simplification below:

\cparagraph{Syntactic program simplification} According to~\cite{nievergelt1965automatic}, syntactic program simplification refers to 
\emph{``simplifications which depend on the form of the program 
only, i.e. can be detected and proved to lead to an equivalent 
program without knowledge of what the program is supposed 
to do''}. This approach focuses on using a set of transformations known to preserve program equivalence. It is aligned with rule-based techniques (such as refactoring) in Table~\ref{tab:Differences}, where equivalent-preserving transformations are applied to enhance program design~\cite{becker1999refactoring}. Therefore, we define \emph{\textbf{syntactic program simplification}} as a method for obtaining simpler programs (with fewer lines of code) through semantic-preserving transformations. Typically, these techniques employ static analysis to verify equivalence.


Refactoring, a representative for syntactic program simplification has been widely studied where most studies focus on the refactoring practice ~\cite{kim2012field,paixao2020behind,peruma2022refactor,alomar2019can}. To reduce the time and effort in manual refactoring, many automated refactoring approaches have been proposed~\cite{gyori2013crossing,binkley2005automated,meng2015does,jensen2010use,karakati2022software}. Several approaches use refactoring to support code review ~\cite{alizadeh2019refbot,brito2021raid,tsantalis2020refactoringminer}. 
Some of the program transformations found in our study are commonly used in refactoring tools (e.g., lambda expressions ~\cite{gyori2013crossing,tsantalis2017clone}). The key difference between refactoring and \simpli{} is that it has the additional criterion that the resulting transformed program should be smaller (reduced lines of code), whereas refactoring does not necessarily lead to reduced code size. 

\cparagraph{Semantic program simplification} Several automatic program simplification techniques \emph{rely on test executions (e.g., test outcome) to check for behavioral equivalence when automatically generating smaller programs} (we call these techniques \emph{\textbf{semantic program simplification}})~\cite{ lu2012automatic,harman1995using,agrawal1990dynamic,ddmin,hammoudi2015use,zhang2013practical,jiang2017simplydroid}. Semantic program simplification have been applied in many domains (e.g., simplifying event sequences in apps~\cite{jiang2017simplydroid}, deobfuscation~\cite{lu2012automatic}).

In terms of transformations used, 
these techniques either rely on deletions (delta debugging) or obtain a reduced list of program statements by focusing on a slicing criterion (dynamic slicing). We refer to these techniques as deletion-based approaches in Table~\ref{tab:Differences}. Several techniques use limited types of transformation (e.g., mutation and crossover operators for pixel shader simplification~\cite{shader}, \emph{remove} and \emph{move}
for debugging concurrent programs~\cite{jalbert2010trace}, replacing identifier and subtrees of parse tree for program reduction~\cite{xu2023pushing}).
These techniques usually rely on a variant of: (1) delta debugging~\cite{ddmin,ddinput}, and (2) dynamic slicing~\cite{agrawal1990dynamic,harman1995using,lu2012automatic,10.1109/ICSE48619.2023.00128,binkley2014orbs}. 
Delta debugging increasingly deletes smaller parts of the input file and run tests to check if the simplified input changes the test outcome~\cite{ddinput,ddmin}.
Dynamic slicing focuses on ``all statements that actually affect the value of a variable 
occurrence for a given program input''~\cite{agrawal1990dynamic}.  Our work is different from dynamic slicing or its variants (e.g., observation-based slicing~\cite{binkley2013observation}) in that: (1) slicing-based techniques aim to isolate feature-related elements (e.g., statements) under certain criteria, whereas we focus on the program itself, we aim to obtain a smaller program (reduced lines of code) which satisfies the criteria; (2) we obtain various transformation types which preserve the original program’s behaviour for program simplification. Our set of transformations serves as the basis for providing more candidates for slicing-based techniques, which facilitate debugging. 

%

\cparagraph{Large language models for program transformation} Large language models (LLMs) have shown promising results in code-related tasks. Most prior learning-based approaches either focus on tasks like method name recommendations~\cite{jiang2019machine,parsa2023method,liu2022learning}, code smell detection~\cite{sharma2021code,liu2019deep,fontana2013code,azeem2019machine,pecorelli2020large} or bug fixing~\cite{zhang2022coditt5,10.1109/ICSE48619.2023.00128}. The most relevant techniques to us are the two general-purpose code transformation models: \tufano{}~\cite{tufano2019learning} and \autotransform{}~\cite{thongtanunam2022autotransform}. Differs from prior approaches, our approach focuses on generating simplified programs with diverse test-equivalent program transformations. 

\cparagraph{\cpsimpli{}} As noted in prior work~\cite{nievergelt1965automatic}, when simplifying a program by hand, developers usually have the required domain knowledge to produce semantically equivalent programs that are smaller. 
Hence, \emph{we study the problem of \simpli{}} that combines the best of both worlds by including (1) syntactic program transformation, and (2) semantic program simplification. 

\begin{definition}[Test equivalent]
\label{def:test}
Given a pair of programs $(P, P')$ and a test suite $T$ executable in $P$ and $P'$, $P$ and $P'$ are test-equivalent if $\forall t_k \in T$  where each test $t_k$ takes as input $inp_k$ and produces output $out_k$, $P$($inp_k$)=$P'$($inp_k$)=$out_k$ (i.e., for all tests in $T$, $P$ and $P'$ produce the same outputs).
\end{definition}

\begin{definition}[Program Simplification]
\label{def:simplification}
Given a program $P$, and a transformed program $P'$, $P'$ is a simplified program of $P$ if (1) $P'$ contains less Source Lines of Code (SLOC) than $P$, and (2) $P$ and $P'$ are test-equivalent. 

\end{definition}
Although there are various granularity levels (e.g., statement, line or character) used in prior techniques, we choose SLOC to represent simplified programs because it has been widely used in deletion-based ~\cite{ddmin,ddinput,hodovan2016modernizing} and rule-based approaches~\cite{dig2009refactoring}. When the SLOCs remain the same, we consider the reduction in the number of tokens.

%
%

\section{Understanding \csimpli{}}\label{sec:study}

\begin{table*}
\centering
\caption{Taxonomy of program simplification in studied OSS projects}
\label{tab:rootcause}
\small
\begin{adjustbox}{width=0.98\textwidth}
\begin{tabular}{lllccc} 
\toprule
\multicolumn{1}{c}{Category}         & \multicolumn{1}{c}{Sub-category}                   & \multicolumn{1}{c}{Description}                                                               & Refactoring & PRs (\#/\%) & Total (\#/\%)                \\ 
\hline\hline
\multirow{10}{*}{(T1) Control logic} & (T1.1) Simplify method return                      & Simplify program logic related to method return value                                         & Y           & 53/13.87    & \multirow{10}{*}{151/39.53}  \\
                                     & (T1.2) Simplify boolean and algebraic expression   & Simplify boolean and algebraic expression by rules                                            & N           & 32/8.42     &                              \\
                                     & (T1.3) Use foreach in loop iteration$^L$           & Foreach loops can~avoid potential off-by-one errors and reduce lines                          & Y           & 15/3.93     &                              \\
                                     & (T1.4) Merge conditional                           & Merge if statements that resulting action is the same                                         & Y           & 13/3.40     &                              \\
                                     & (T1.5) Ternary conditional operator$^L$            & Simplify conditional statements via ternary conditional operator                              & Y           & 11/2.88     &                              \\
                                     & (T1.6) Restructure conditional branches            & Replace complex conditional branches by using enum, polymorphism, and etc.                    & Y           & 9/2.36      &                              \\
                                     & (T1.7) Replace with pipeline                       & Chain operations together by stream pipelining                                                & Y           & 6/1.57      &                              \\
                                     & (T1.8) Replace variable with attribute             & Replace variables in method with attributes                                                   & Y           & 6/1.57      &                              \\
                                     & (T1.9) Merge catch ~~                              & Merge multiple catch together if they contain duplicated code                                 & Y           & 5/1.31      &                              \\
                                     & (T1.10) Change return type                         & Change method return to void and delete statements                                            & Y           & 1/0.26      &                              \\ 
\hline
\multirow{3}{*}{(T2) Extraction}     & (T2.1) Extract method                              & Extract code blocks as methods to improve reusability                                         & Y           & 75/19.63    & \multirow{3}{*}{120/31.41}   \\
                                     & (T2.2) Extract variable                            & Extract common variables                                                                      & Y           & 40/10.47    &                              \\
                                     & (T2.3) Consolidate duplicate conditional fragments & Pull up common head or pull down common tail of conditional to reduce duplication             & Y           & 5/1.31      &                              \\ 
\hline
\multirow{3}{*}{(T3) Deletion}       & (T3.1) Remove~unnecessary~code                     & Remove duplicated or unneeded code                                                            & N           & 47/12.30    & \multirow{3}{*}{63/16.49}    \\
                                     & (T3.2) Remove unused imports                       & Remove unused imports                                                                         & N$^T$       & 11/2.88     &                              \\
                                     & (T3.3) Clean up dead code blocks                   & Remove dead code blocks that can not be visited                                               & Y           & 5/1.31      &                              \\ 
\hline
(T4) API                             & (T4.1)~Replace with equivalent API                 & Replace code block with semantically-equivalent APIs                                          & N           & 62/16.23    & 62/16.23                     \\ 
\hline
\multirow{2}{*}{(T5) Inline code}    & (T5.1) Inline variable                             & Inline temporary variables that are only used once                                            & Y           & 38/9.95     & \multirow{2}{*}{47/12.30}    \\
                                     & (T5.2) Inline method                               & Inline simple small method that are only used once                                            & Y           & 9/2.36      &                              \\ 
\hline
(T6) Lambda                          & (T6.1) Use lambda$^L$                              & Simplify using lambda expression                                                              & Y           & 44/11.52    & 44/11.52                     \\ 
\hline
\multirow{6}{*}{(T7) Others}         & (T7.1) Use diamond operator$^L$                    & Simplify instantiation of generic class (since Java 1.7)                                      & N$^T$       & 16/4.19     & \multirow{6}{*}{46/12.04}    \\
                                     & (T7.2) Code style reformat                         & Use concise code style                                                                        & N$^T$       & 9/2.36      &                              \\
                                     & (T7.3) Use constructor to initialize               & Use class constructor to initialize the class properties                                      & Y           & 7/1.83      &                              \\
                                     & (T7.4) Merge imports                               & Merge multiple imports from the same package                                                  & N$^T$       & 6/1.57      &                              \\
                                     & (T7.5) Replace with annotations$^L$                & Replace code fragments with annotations which achieve the same functionality                  & Y           & 6/1.57      &                              \\
                                     & (T7.6) Try-with-resources$^L$                      & Use try-with-resources statement instead of finally block to close resources (since Java 1.7) & N$^T$       & 2/0.52      &                              \\
\bottomrule
\end{tabular}
\end{adjustbox}
\begin{tablenotes}
\footnotesize
\item{L} The superscript `L' indicates the program transformation uses language-specific feature (e.g., foreach, lambda, diamond operator)
\item{T} The superscript `T' indicates that although it is not a refactoring type, there is IDE plugin that supports the transformation.
\end{tablenotes}
\end{table*}

We study \simpli{} by manually inspecting commits within pull requests (PRs) in Java open-source GitHub projects. We choose Java because it is one of the most popular programming languages. 
We concentrate on PRs as they contain detailed discussions to help us classify the simplification types and understand the motivations behind them.

\subsection{Research Methodology}
\label{sec:studymet}
\subsubsection{Mine \simpli{} in PRs}
We developed a crawler using the GitHub API to search for PRs related to code simplification using keyword ``simplify code''~\footnote{https://developer.github.com/v3/}. We tried other keywords like ``shorten'', ``reduce'', and ``shrink'' and found out that the resulting PRs contained too many non-simplified PRs. 
After retrieving the top 1000 relevant PRs, we manually reviewed each one to exclude those that did not clearly focus on simplification, were irrelevant to code simplification, or did not reduce lines of code. This process resulted in \analyzedPRs{} PRs from \analyzedRepos{} repositories.

\subsubsection{Derive taxonomy of transformation types} 

After receiving simplification commits in PRs, we developed a taxonomy through manual analysis of code changes in each PR using thematic analysis~\cite{cruzes2011recommended} -- an approach that identifies patterns (or ``themes'') within data. To this end, we recruited two human raters to develop the taxonomy. Both raters were graduate students with over two years of Java programming experience. They followed the following steps independently, with conflicts resolved in meetings. First, we carefully reviewed PR titles, descriptions, and discussions to understand developers' motivations and the simplification process. We identified simplification commits by examining commit messages. If no explicit simplification-related keywords were found, we checked all PR commits.
Next, we coded key ``diff hunks" (i.e., code changes performing simplification) in each commit, describing transformation types. We iteratively refined codes by reviewing related diff hunks and their context. We grouped key diff hunks based on codes, providing an overview of main edit actions and recurring patterns. Codes with similar meanings were aggregated into groups to derive broader themes. Finally, we reviewed and merged similar themes to define the final set, reducing research bias.  During the rating, our rators achieved an initial Cohen's Kappa of \studycohenkappa{}. Disagreements were resolved, resulting in a Cohen's Kappa of 1.0.

\subsection{RQ1: Set of Transformations}
\label{sec:pattern}
Before following the steps of thematic analysis, we reviewed syntactic and semantic program simplification principles together with program refactoring from prior studies~\cite{nievergelt1965automatic,ddinput}. 
Table~\ref{tab:rootcause} lists the identified types of program transformations across \analyzedPRs{} PRs across \analyzedRepos{} projects used in \simpli{}. Note that a PR may contain multiple diff hunks where each diff hunk group may correspond to a particular transformation type, leading to multiple transformation types. The ``Category''
 column in Table~\ref{tab:rootcause} describes the high-level types of the program simplification, while the
``Sub-category'' column gives the specific categories. The last column (Category Total (\#/\%)) presents the total number and percentage of PRs that fit into a particular category. In total, we identified \numberofmaincategories{} main categories with \numberofsubcategories{} sub-categories. Among the \numberofmaincategories{} categories, we observed that mere deletion only exists in \delcatstudy{}\% of the investigated PRs. This indicates that there \emph{exists a missing gap in prior deletion-based approaches} (marked as ``D'' in Table~\ref{tab:Differences}), 
i.e., they can only fulfil the needs of up to \delcatstudy{}\% \simpli{}.  

``Extract method'' is the most commonly used transformation. This result is inline with prior studies that revealed that ``extract method'' is the most popular and most well-motivated refactoring among developers~\cite{silva2016we}. 
As our study shows that ``extract method'' is frequently used in \simpli{}, it provides empirical evidence that developers still often perform ``extract method'' refactoring manually. However, prior study shows that manual refactoring often leads to mistakes~\cite{murphy2008breaking}. 

Table~\ref{tab:rootcause} also shows that there are \emph{simplification types that are Java language-specific} (foreach, ternary conditional operator, lambda, diamond operator, and annotations). This indicates that developers prefer using language features to simplify programs. Meanwhile, designers of future automated program simplification tools should incorporate these language features. 

\begin{tcolorbox}[left=1pt,right=1pt,top=1pt,bottom=1pt]
\vspace{-4pt}
\textbf{Finding 1:} \cpsimpli{} includes \numberofsubcategories{} transformation types where deletion only exists in \delcatstudy{}\% of studied PRs. ``Extract method'' (\extractmethodratio{}\%) and \useapi (\apiratio{}\%) are the two most frequently used  transformations. Some transformations involve language-specific features.  

\textbf{Implication 1:} The types of transformations used in \simpli{} are diverse. To enhance reusability, developers often introduce new method or replace with an API call. When simplifying programs, developers also often use language-specific features (e.g., foreach, lambda). 
Several widely used transformations (e.g., \useapi) have not been automated by any existing tool, indicating a need to design new tools to support these transformations.
\vspace{-3pt}
\end{tcolorbox}

\subsection{RQ2: Motivations Behind Simplifications}
\label{sec:motivation}

We carefully read each PR's title, description and discussions to analyze the motivations for simplification. In some PRs, developers only state ``what'' (changes they have done) without ``why'' (the motivations), so for RQ2, we only keep PRs with clear descriptions of the motivations. We use the same procedures described in Section~\ref{sec:studymet}. This results in \analyzedMotivationPRs{} out of total \analyzedPRs{} studied PRs.

\vspace{-0.4cm}
\begin{table}[H]
\centering
\caption{Motivations for simplifying programs}
\label{tab:motivations}
\footnotesize
\begin{tabular}{llr} 
\toprule
\textbf{Category}    & \textbf{Description}             & \multicolumn{1}{l}{\textbf{GitHub PRs (\#/\%)}}  \\ 
\midrule
Cleanup code~\cite{ddmin,ddinput,hodovan2016modernizing}    & Remove unnecessary code & 49/62.03                                \\ 
Readability~\cite{buse2008metric,posnett2011simpler,munoz2020empirical} & Improve readability & 21/26.58                                \\ 
Complexity~\cite{mccabe1976complexity,campbell2018cognitive,ebert2016cyclomatic}  & Reduce complexity   & 7/8.86                                 \\ 
Reusability~\cite{bansiya2002hierarchical,washizaki2004metrics,rathee2022metrics}
& Reuse existing code     & 3/3.80                                 \\
\bottomrule
\end{tabular}
\end{table}
\vspace{-0.5cm}

Table~\ref{tab:motivations} shows the \numberofmotivations{} main motivations, including ``Readability'' (\readabilityratio{}\%), ``Complexity'' (\complexityratio{}\%), ``Cleaning up code'' (\deletionratio{}\%), and ``Reusability'' (\reusabilityratio{}\%). Note that one PR might contain multiple motivations for program simplification. Compared to prior refactoring studies~\cite{silva2016we,alomar2021we,kim2012field}, the main motivations of \simpli{} are generally similar, with the focus being on removing duplication. Notably, we also observe from Table~\ref{tab:motivations} that \emph{all motivations of \simpli{} can be automatically measured using prior metrics}, indicating the promise of using existing metrics to 
assess the quality of simplified programs. 


\begin{tcolorbox}[left=1pt,right=1pt,top=1pt,bottom=1pt]
\vspace{-3pt}
\textbf{Finding 2:} The motivations that drive \simpli{} is generally similar to those of refactoring. These include: (1) cleaning up code, (2) improving readability, (3) reducing complexity, and (4) improving reusability. All motivations can be automatically measured using prior code metrics. 

\textbf{Implication 2:} 
As all motivations can be automatically measured using prior metrics, this shows the promise of using these metrics to assess the quality of simplified programs.
\vspace{-3pt}
\end{tcolorbox}


\subsection{Supported Refactoring Types}
\label{sec:overlapping}

We conducted two studies to analyze the overlap in program transformations between \simpli{} and those supported by refactoring tools. First, we used RefactoringMiner~\cite{tsantalis2020refactoringminer,Tsantalis:ICSE:2018:RefactoringMiner} to identify if our transformations correspond to known refactorings, supplemented by prior studies and Martin Fowler’s refactoring catalog~\cite{becker1999refactoring} for identifying refactoring-based program simplifications. Second, to understand the detection challenges, we ran RefactoringMiner on simplification commits and contacted its developers to confirm if each transformation fits into currently supported refactoring types, understand detection challenges, and report detection failures.

\subsubsection{\textbf{Covered by refactoring}}
Table~\ref{tab:rootcause} shows the overlapping (``Y'') and non-overlapping transformations (``N'') with refactoring types and those supported as IDE plugins (``T'').
 Notably, control-logic modifications (\controllogicrelatedratio{}\%) are frequently seen in \simpli{}, but refactoring tools only support a subset. Additionally, \extractionratio{}\% of the \simpli{} involves extraction, a prevalent refactoring technique replacing code blocks with equivalent methods, variables, or statements.

\subsubsection{\textbf{Beyond refactoring}}
Some simplifications do not align with well-established refactoring types. The second-frequently used type (\apiratio{}\%) replaces method code snippets with semantically equivalent APIs, like \texttt{Arrays.fill()} or \texttt{StringUtils.isEmpty()}. Listing~\ref{lst:apis_example} gives an example for using \texttt{Arrays.fill()} to replace the code block that implements the same function. While existing studies on API recommendation~\cite{huang2018api,rahman2016rack,he2021pyart} focus on suggesting APIs based on language descriptions, we found no technique that automatically substitutes code blocks with equivalent API calls to improve robustness readability and reduce code size.

\begin{lstlisting}[style=mystyle, escapechar=^,caption=Example of \useapi,upquote=true,label={lst:apis_example}]
- Entry[] tab = table;
- for (int i = 0; i < tab.length; i++)
-   tab[i] = null;
+ Arrays.fill(table, null);
\end{lstlisting}

Based on the discussion with RefactoringMiner's developers, \useapi{} can be theoretically supported via combined operations that perform multi-project code clone detection, aggregate similar code snippets, extract methods and then invoke these methods. Moreover, when selecting the appropriate APIs to replace, automated tools need to consider (1) the security of the used APIs~\cite{madden2020api} and security of the dependencies of the used APIs and (2) API misuse~\cite{zhang2018code,nguyen2019api} as inexperienced programmers may use API incorrectly.

\noindent\emph{Simplify boolean expression (\#\conditionalexpressionnumber{}, \conditionalexpressionratio{}\%).}  Simplifying boolean expressions can lead to cleaner code. 
Listing~\ref{lst:conditional_example} shows an example where the expression \texttt{false == co.isExpired()} is simplified via the semantically-equivalent expression \texttt{!co.isExpired()}. 
Moreover, simplification can preserve errors and non-termination (i.e., if \texttt{false == co.isExpired()} throws an exception or gets stuck in an infinite loop, the simplified expression still has the same behavior).  In principle, we can simplify algebraic expressions by using rules like cancellation, commutativity, associativity, and design rules that enumerate all such semantically-equivalent expressions. However, it is time-consuming to craft these rules and infeasible to include all equivalent expressions. 
\begin{lstlisting}[style=mystyle, escapechar=^,caption=Example of ``Simplify boolean expression'',upquote=true,label={lst:conditional_example}]
- if (false == co.isExpired()) {
+ if (!co.isExpired()) {
    return true;}
\end{lstlisting}


\noindent\emph{Diamond operator (\#\diamondnumber{}, \diamondratio{}\%).} The diamond operator has been introduced since Java 1.7 to simplify genetic instantiation. Listing~\ref{lst:diamondexample} shows an example of using the diamond operator for simplification. 
\begin{lstlisting}[style=mystyle, escapechar=^,caption=Example of ``using diamond operator",upquote=true,label={lst:diamondexample}]
- Set<String> conditionKeys = new HashSet<String>();
+ Set<String> conditionKeys = new HashSet<>();
\end{lstlisting}

\noindent\emph{Try-with-resources (\#\trywithnumber{}, \trywithratio{}\%).} The try-with-resources statement is a try statement that declares one or more resources in it. A resource is an object that must be closed once your program is done using it. For example, a File resource or a Socket connection resource. The try-with-resources statement ensures that each resource is closed at the end of the statement execution. So the programmer don't have to manually close those resources in the \texttt{finally} block. Listing~\ref{lst:trywithexample} shows an example of using the diamond operator for simplification. 
\begin{lstlisting}[style=mystyle, escapechar=^,caption=Example of ``Try-with-resources",upquote=true,label={lst:trywithexample}]
- pluginManager.lock();
- try {
+ try (ResourceLock ignore = pluginManager.obtain()){
    ...
- finally {  pluginManager.unlock(); }
\end{lstlisting}

\begin{tcolorbox}[left=1pt,right=1pt,top=1pt,bottom=1pt]
\vspace{-3pt}
\textbf{Finding 3:} While most simplifications (70\%) are covered by existing refactoring types, 30\% require new transformations: \useapi{} (17\%), ``Remove unnecessary code '' (13\%), ``Simplify boolean expression'' (9\%), ``Use diamond operator'' (4\%), ``Remove unused imports'' (3\%), and ``Try-with-resources'' (0.5\%). Furthermore, five transformations are supported as IDE plugins.

\textbf{Implication 3:} 
Some newly discovered transformations are available as standalone IDE plugins but cannot be found under the refactoring menu. As IDE users have expressed difficulty in finding these transformations~\cite{organize,diamond}, it is worthwhile to group together program simplifications to enhance usability. Meanwhile, most newly discovered \simpli{} cannot be easily supported by existing refactoring engines due to (1) missing API information (\useapi{}), (2) a lack of exhaustive semantic-equivalent rules (`Simplify boolean expression'' and ``Simplify non-control statement''),  and (3) uncertainty about redundant code (``Remove unnecessary code ''). This indicates the need to design new program simplification tools to automate these transformations.
\vspace{-3pt}
\end{tcolorbox}

For each commit, we also run \refactoringminer{} to check if it can detect the transformation. Notably, we found several scenarios where \refactoringminer{} fails to detect the transformations despite being in the list of supported refactoring types. 
Listing~\ref{lst:simplifyreturnexample}~\cite{inlinevariableexamplecommitlink} shows an example where the ``Inline Variable'' refactoring interleaved with method return operation, which we classified as `` Simplify method return'' (\functionreturnratio{}\%). Although \refactoringminer{} supports this transformation type, its newest version (V2.0.4 chrome extension) failed to detect the example. After analyzing the implementation of \refactoringminer{}, we notice that 
the commit where the
``Inline Variable'' refactoring is interleaved with the return statement
causes the human-crafted rules of \refactoringminer{}  to fail.

\begin{lstlisting}[style=mystyle, escapechar=^,caption=\refactoringminer{} fails to detect ``Simplify method return'' due to interleaved transformations, upquote=true, label={lst:simplifyreturnexample}]
  public String create() {
-   String token = UUID.randomUUID().toString();
-   return token;
+   return UUID.randomUUID().toString();
  }
  private String createToken() {
    CsrfToken csrfToken = new CsrfToken();
-   String token = csrfToken.create();
-   return token;
+   return csrfToken.create();
  }
\end{lstlisting}


We also found cases where \refactoringminer{} cannot detect the complete mapping for a multi-line transformation. For example,
although \refactoringminer{} can detect the addition, removal, and modification of the annotations by matching annotation type between two commits~\cite{kim2021studying}, it fails to provide mapping 
to the users if they want to know which part of the code is replaced by the added annotations. 
Listing~\ref{lst:annotationexample} shows an example for the methods \texttt{getDefaultName()} and \texttt{setDefaultName(String defaultName)} that are replaced with the annotations \texttt{@Getter} and \texttt{@Setter}. \refactoringminer{} can only detect that the two annotations are added but fails to report the relationship between the added annotations with the getter and setter. \refactoringminer{} also offers limited support for transformations with language-specific features. Except for replacing anonymous with lambda, other simplifications using language-specific features (e.g., foreach, diamond operator, try-with-resources) cannot be detected by \refactoringminer{} (\languagefeatureratio{}\% of studied PRs). 
\begin{lstlisting}[style=mystyle, escapechar=^,caption=Example of  ``Replace with annotations'', upquote=true, label={lst:annotationexample}]
+ @Getter
+ @Setter
public class TestAppConfiguration {  
private String defaultName;
- public String getDefaultName() {
-   return defaultName;
- }
- public void setDefaultName(String defaultName) {
-   this.defaultName = defaultName;
- }
}
\end{lstlisting}

\begin{tcolorbox}[left=1pt,right=1pt,top=1pt,bottom=1pt]
\vspace{-3pt}
\textbf{Finding 4:} Among the supported refactoring types, \refactoringminer{} is still limited in: (1) handling refactorings that are not pure (interleaved with other transformations), (2) providing a complete mapping for a multiline transformation, and (3) handling language-specific features. 
These limitations requires designing the rules for detection in \refactoringminer{}.

\textbf{Implication 4:} Our study reveals several limitations of \refactoringminer{}, including (1) the detection rules for current refactoring should consider the complex situation with interleaved transformations, (2) the detection for annotation should be more fine-grained, the refactoring engine should offer maps between the replaced code snippet with changed annotations, not only show which annotations are added, deleted, or modified, (3) more support should be offered to the transformations related with language-specific feature, considering \languagefeatureratio{}\% of our studied PRs are related with it, and its prevalence in the OSS repositories.
\vspace{-3pt}
\end{tcolorbox}

\subsubsection{\textbf{Automation}}
\label{sec:auto}
We analyze the possibility of automating program simplification by analyzing the supported simplifications by two auto-refactoring tools~\cite{tsantalis2018ten,10.1145/2932631}.
\jdeo{}~\cite{tsantalis2018ten} is a code smell detection tool that automatically identifies and applies refactoring. \jdeo{} supports five code smells (Feature Envy, Type/State Checking, Long Method, God Class and Duplicated Code) where ``Long Method'' can be resolved by ``Extract Method'' refactoring, and ``Duplicated Code'' problems can be resolved by ``Extract Clone'' refactoring. 
As our study shows that only \extractionratio{}\% simplifications are related to the extraction, only up to \extractionratio{}\% simplifications can be recommended by \jdeo{} in the best case. Meanwhile, prior multi-objective approach supports 11 refactoring types~\cite{10.1145/2932631} where three are simplification types. 

After carefully analyzing studied simplifications, we identify the following features: (1) the program token length before and after simplification are relatively short (within 512 tokens). For example, the program before simplifying using \useapi{} is usually a contiguous code snippet within a method scope, whereas the simplified program only contains a few lines of API invocations; (2) the simplification types have some recurrent patterns; (3) massive simplification data can be mined from GitHub. Based on these features, we believe that deep-learning-based technology can be used to automate program simplification. 

\begin{tcolorbox}[left=1pt,right=1pt,top=1pt,bottom=1pt]
\vspace{-3pt}
\textbf{Finding 5:} Auto-refactoring tools (like \jdeo{}) can be used to automate \simpli{} but they only support up to seven of \numberofsubcategories{} simplification types. 

\textbf{Implication 5:} Prior auto-refactoring tools fail to cover most simplification types. Using deep-learning-based techniques trained on a richer set of transformations could be promising. 
\vspace{-3pt}
\end{tcolorbox}



\section{Methodology}

\begin{figure}[t]
\centering
\includegraphics[width=\linewidth]
{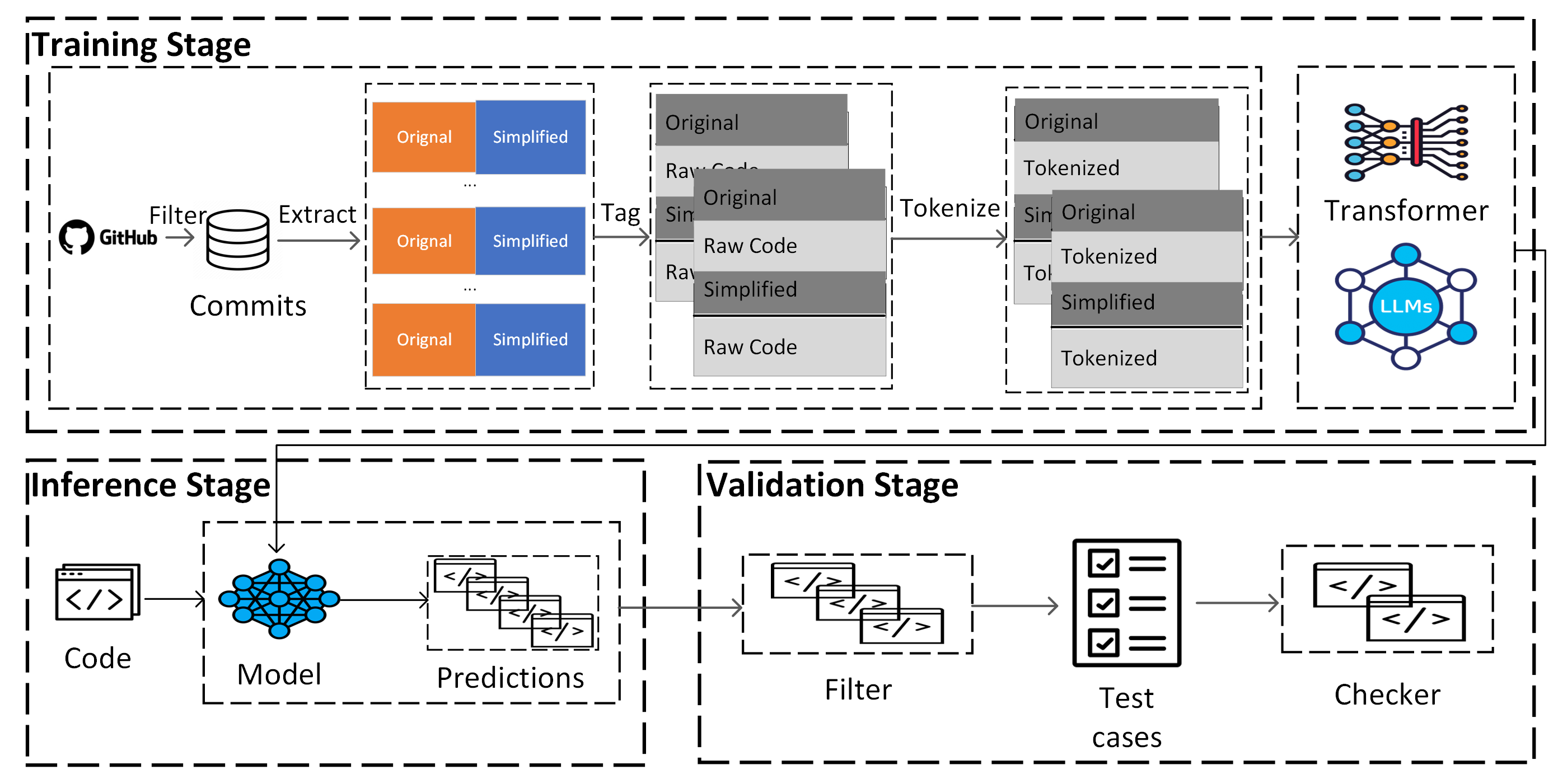}
\caption{Overall workflow of \tooln.}
\centering
\label{Figure:overview}
\end{figure}


Based on our study, we design \tooln{}, a deep learning-based program simplification framework.
Figure~\ref{Figure:overview} depicts the overall workflow of \tooln{} including three stages: (1) training, (2) inference, and (3) validation. 



\cparagraph{Training and test data collection} Although there are several refactoring datasets available~\cite{kadar2016code,hegedHus2018empirical}, some 
simplification types are not covered by existing refactoring.  
 Hence, we propose \bench, a new program simplification dataset that contains pairs of ($P$, $P'$) with the original method $P$ and its simplified version $P'$.
\begin{table}[t!]
\centering
\caption{Statistics of \bench}
\label{tab:statistics}
\small
\begin{tabular}{l|rr} 
\toprule
Statistic                         & \multicolumn{1}{l}{Whole} & \multicolumn{1}{l}{Valid}  \\ 
\midrule
\# of Projects                    & 25,022                     & 307                        \\
\# of Method-pairs                & 92,485                     & 404                        \\
Mean \# of tests in projects      & -                         & 96                         \\
Mean SLOCs \# of original method   & 17                        & 16                         \\
Mean SLOCs \# of simplified method &  14                         &  12                          \\
\bottomrule
\end{tabular}
\end{table}

\noindent\textbf{Select \csimpli{} Commits.} We use GHTorrent~\cite{gousios2012ghtorrent} to collect the related commits between 2011 and 2021. To remove commits irrelevant to program simplification, we only keep commits with the keywords ``simplify'' or its derived words ``simplification'', ``simplified'', together with ``code'' or ``program'' in their commit messages. As we only focus on simplification for Java programs, we also filter commits where the committed code changes contain no Java files. This results in \bigdatasetcommitnumber{} commits from \bigdatasetreponumber{} projects. To understand the noise this filtering mechanism can potentially introduce, we manually examined a randomly selected sample of 100 commits filtered using this approach and found 89 simplification-related commits. The manual analysis shows that the amount of noise (11\%) is reasonable for ML training on large training data~\cite{kim2011dealing}. Next, we split a commit according to \emph{diff hunks}. As stated in Definition~\ref{def:simplification}, program simplification aims to reduce SLOCs while preserving the test-equivalent behavior of the generated program. Hence, our crawler automatically selects diff hunks which contain more deleted lines than added lines (code comments and blank lines in the diff hunk are excluded while counting the lines of code). We extract the method surrounding the diff hunk because (1) it contains a logical set of relevant variables and could give semantic information about the functionality~\cite{tufano2019empirical,lutellier2020coconut, jiang2021cure}; (2) compared to class, the method is relatively short with fewer tokens, which is more friendly for deep learning model training or tuning as maximum input and output sequence length is usually limited (e.g., the maximum input sequence lengths for CodeT5 is 512~\cite{wang2021codet5}). In total, our collected dataset contains \totnumpairbench{} (original method, simplified method) pairs from \bigdatasetreponumber{} projects.

\noindent\textbf{Data Preprocessing.}
After collecting the code samples, we then pre-process the raw source into pairs at the method level, where each training sample contains the original code and the simplified version~\cite{10.1109/ICSE48619.2023.00128,le2012genprog,lutellier2020coconut,zhu2021syntax,liu2019tbar}.
As shown in Figure~\ref{Figure:overview}, our code representation is based on the token sequence as prior large language models trained on code (LLMC). Like~\cite{tufano2019empirical,lutellier2020coconut, jiang2021cure}, \tooln{} takes the tokenized original method as input and produces the simplified method as output. We use the byte-pair encoding (BPE) scheme ~\cite{sennrich2015neural} to build our tokenizer.
To avoid data leakage, we also test the trained model on code samples extracted from \emph{new, unseen} projects. In our evaluation, we split the data into training, validation, and testing sets with a ratio of 8:1:1 at the project level (80\% of projects into training, 10\% into validation, and 10\% into testing).

 Table~\ref{tab:statistics} shows the statistics of \bench. As some unmodified versions of projects in our dataset may have syntax errors or test failures, we divide the dataset into (1) \emph{whole} (include all collected simplified method pairs) and (2) \emph{valid} (a subset of the whole dataset where the unmodified projects can be compiled successfully and all tests within the test suite of the project pass). To prepare the \emph{valid} dataset, we filter commits that (1) fail to compile using \texttt{mvn build} command, and (2) do not have any test. Table~\ref{tab:statistics} shows that projects in \bench{} contain a relatively large number of tests (on average \validaveragetestnumber{} tests), which allows us to check for test-equivalent. 

\subsection{Model Training and Tuning} 

\cparagraph{Simplified line localization}  
Inspired by APR approaches~\cite{10.1109/ICSE48619.2023.00128,le2012genprog,lutellier2020coconut,zhu2021syntax,liu2019tbar} where fault localization is performed to identify the ``buggy'' lines in which the transformations are applied, we propose \emph{simplified line localization} to pinpoint the lines in the original program $P$  to apply the program transformations from \simpli{} to produce the simplified program $P'$.
To encode simplified lines into our code representation, we marked the original and the simplified code lines with special tokens (<original>, </original>, <simplified>, </simplified>). 
If multiple \emph{diff} hunks exist, we will add these tokens to each changed line at the beginning and end. We evaluate the effect of \emph{perfect simplified line localization} (i.e., the lines modified by the correct \simpli{}) for program simplification, following prior APR evaluation~\cite{lutellier2020coconut,zhu2021syntax}. 
To evaluate the impact of localization information on program simplification, we also design another code representation without simplified line localization (i.e., with no special tokens inserted) as a baseline comparison. In Section~\ref{sec:localizeresults}, our experiment shows that adding simplified line localization helps improve the model's effectiveness.

\cparagraph{Model design} Given the nature of our task, we use an Encoder-Decoder-based LLMC~\cite{wang2021codet5,fu2022vulrepair} as it is shown to be effective on code transformation tasks. Our LLMC is trained through training samples and  hard prompts~\cite{gu2021ppt,wang2022no}. By adding fixed natural language instructions to the model input, hard prompt uses task-specific knowledge learned during pre-training for the subsequent tuning stage. 
Specifically, we use the prompt
``\emph{Simplify the following java method: [X], the simplified version is: [Y]}'' where [X] represents the original method and [Y] represents the simplified method. This natural language instruction is designed for CodeT5.

\cparagraph{Simplified program generation}
In the generation phase, we use a common search strategy called beam search to generate and rank a large number of candidate predictions. For each iteration, the beam search algorithm checks the $n$ most likely tokens (n corresponds to the beam width) and ranks them by the total likelihood score of the next $s$ prediction steps (s corresponds to the search depth). Subsequently, the beam search algorithm outputs the top $t$ most likely sequences ordered based on the likelihood of each sequence.

\cparagraph{Filtering unaltered programs} Among the generated simplified programs, we notice that about \unchangeditemratio{}\% of them are the same with the original method (we call these programs \emph{unaltered programs}). As unaltered programs are naturally semantically equivalent to the original programs and may mislead the validation stage, we filter out all unaltered programs before the validation.

\subsection{Validation Stage}
Validation is the key step in ensuring the quality of the generated simplified programs. As stated in
Definition~\ref{def:simplification}, given an original program $P$, we consider a program as a simplified program $P'$ if (1) $P'$ contains less SLOCs than $P$, and (2) $P'$ is test-equivalent with $P$. To check for the (1) condition, we compare the SLOCs of $P$ and $P'$ to ensure that the code size of $P'$ is reduced. To check for the (2) condition, we first filter out simplified programs that do not compile and then run each simplified program against the entire test suite to check for test equivalent. Based on Finding 2, we designed a checker to measure the differences between the original and simplified programs in complexity and readability.

\cparagraph{Cyclomatic complexity}  One of the main motivations for \simpli{} is to reduce the complexity of a program. We use cyclomatic complexity~\cite{mccabe1976complexity,campbell2018cognitive,ebert2016cyclomatic}, a widely used metric to measure the complexity of a program.

\cparagraph{Readability} We use cognitive complexity ~\cite{munoz2020empirical} to measure the readability of the simplified program. Cognitive complexity measures how difficult it is for humans to read and understand a method.

\section{Evaluation}
\label{sec:evaluation}
We evaluate the effectiveness of \tooln in automated program simplification and answer the following research questions: 
\begin{description}[leftmargin=*]
\item[RQ5] How effective is \tooln ~ compared to other techniques? 
\item[RQ6] How effective is 
simplified line localization in \tooln{}?

\item[RQ7] Among the correctly generated programs, what are the transformation types used to generate the simplified programs? 
\item[RQ8] What is quality of the generated programs by \tooln{} compared to other tools?
\end{description}

\subsection{Experimental Setup}


\subsubsection{Implementation} We use the pre-trained CodeT5-base model and corresponding tokenizer in the official repository Huggingface~\cite{Huggingface}. 
Our implementation of prompt-tuning is based on OpenPrompt~\cite{OpenPrompt}. We use the generic training strategy and parameter settings following the official implementation of CodeT5~\cite{wang2021codet5}. Specifically, we set the learning rate to be $5e^{-5}$ and use AdamW optimizer with a linear warmup to optimize the model. The training and validation batch size is 8. We tun the CodeT5-base model for 50 epochs. The maximum length of input and output text is set to 512. During training, we set beam size to 1 and 10. All experiments are conducted on Ubuntu 20.04 with 2x 24GB GeForce RTX 3090 GPUs. For the generated simplified programs, we first replace the original method with the simplified one, then compile the entire project under JDK1.8.0 and JDK11.0.15 during the validation step. When validating candidate programs, we sample from the model up to beam size = 1 and 10 predictions to be validated. \tooln{} validates simplifications candidates automatically where only the first one (if any) is found to be test-equivalent will be provided to users. 

\subsubsection{Evaluation metrics} We use \emph{prefect prediction}~\cite{tufano2019learning, thongtanunam2022autotransform} and CodeBLEU
~\cite{ren2020codebleu}  to evaluate the effectiveness of each learning-based baseline. The first metric quantifies the ratio of generated programs that match the test samples' ground truth (i.e., the developer-implemented after-simplification version). The second metric measures the similarity of the generated code to the ground truth. Prior study~\cite{ren2020codebleu} showed that the CodeBLEU metric better correlates with developers’ perception of code similarity than the BLEU metric. Both metrics are commonly used by learning-based approaches to assess the quality of the generated programs.

\subsubsection{Baselines}
We compare \tooln with the following approaches:

\cparagraph{Ideal Delta debugging (IDD)} As there many variants of delta debugging~\cite{wang2021probabilistic,ddinput,misherghi2006hdd}, we adopt the ideal version of delta debugging IDD as a strong representative of these variants.  Specifically, IDD (1) will only use deletions for transformations, and (2) can select the \emph{correct} statements for deletions (whenever simplifying the ground truth $G$ requires only pure deletions, IDD can produce exactly the same program as $G$). 
IDD represents deletion-based approaches.

\cparagraph{\jdeo} This is auto-refactoring tool~\cite{tsantalis2018ten} introduced in Section~\ref{sec:auto}. We use \jdeo to represent rule-based approaches.


\cparagraph{\tufano} This is an NMT-based sequence-to-sequence transformation model~\cite{tufano2019learning} . It tokenizes the input program into a sequence and translates the sequence to a fixed code sequence.

\cparagraph{\autotransform} This method leverages a BPE scheme to handle new tokens and a Transformer-based NMT architecture to handle longer sequences~\cite{thongtanunam2022autotransform}.

\cparagraph{Vanilla transformer} As we are the first to study the program simplification task, and other baselines are not specifically designed for this task, we also train a vanilla transformer from scratch on our whole dataset by reusing the parameter setting in prior work~\cite{thongtanunam2022autotransform}.

\subsection{[RQ5] Comparison with Prior Techniques}
\label{sec:evaltools}

\begin{table*}
\centering
\caption{Experimental results of deep-learning-based baselines}
\label{tab:results_of_deep_learning_baselines}
\begin{adjustbox}{width=0.90\textwidth}
\begin{tabular}{l|l|r|rrr|r|rrr|r|rrr|r|rrr} 
\toprule
\multirow{3}{*}{Benchmark}                                               & \multirow{3}{*}{Code Rep.} & \multicolumn{4}{c|}{\tufano}             & \multicolumn{4}{c|}{\autotransform}                      & \multicolumn{4}{c|}{Vanilla Transformer}   & \multicolumn{4}{c}{\tooln}                                                                                                                \\ 
\cline{3-18}
                                                                         &                            & PP      & \multicolumn{3}{c|}{CodeBLEU} & \multicolumn{1}{l|}{PP} & \multicolumn{3}{c|}{CodeBLEU} & PP         & \multicolumn{3}{c|}{CodeBLEU} & \multicolumn{1}{c}{PP} & \multicolumn{3}{c}{CodeBLEU}                                                                                     \\
                                                                         &                            & \#/\%   & mean  & median & st. dev.     & \#/\%                   & mean  & median & st. dev.     & \#/\%      & mean  & median & st. dev.     & \#/\%                  & mean                                & median                              & st. dev.                             \\ 
\hline
\multirow{2}{*}{\begin{tabular}[c]{@{}c@{}}Whole\\(\#9508)\end{tabular}} & Raw                        & 10/0.11 & 0.450 & 0.423  & 0.205        & 0/0.00                  & 0.094 & 0.060  & 0.093        & 335/3.52   & 0.801 & 0.828  & 0.148        & \textbf{1842/19.37}    & \textbf{0.856}                               & \textbf{0.882}                               & 0.136                                \\ 
\cline{2-18}
                                                                         & Localized                  & 14/0.15 & 0.411 & 0.385  & 0.199        & 0/0.00                  & 0.084 & 0.053  & 0.088        & 2738/28.80 & 0.867 & \textbf{0.911}  & 0.149        & \textbf{2827/29.73}    & \textbf{0.878}                               & \textbf{0.911}                               & 0.128                                \\ 
\hline
\multirow{2}{*}{\begin{tabular}[c]{@{}c@{}}Valid\\(\#404)\end{tabular}}  & Raw                        & 0/0.00  & 0.460 & 0.449  & 0.195        & 0/0.00                  & 0.086 & 0.053  & 0.089        & 2/0.50     & 0.317 & 0.302  & 0.172        & \textbf{64/15.84}      & \textbf{0.849} & \textbf{0.873} & 0.132 \\ 
\cline{2-18}
                                                                         & Localized                  & 0/0.00  & 0.425 & 0.397  & 0.196        & 0/0.00                  & 0.086 & 0.052  & 0.092        & 2/0.50     & 0.386 & 0.396  & 0.199        & \textbf{123/30.45}     & \textcolor[rgb]{0.2,0.2,0.2}{\textbf{0.880}} & \textbf{0.910} & 0.124  \\
\bottomrule
\end{tabular}
\end{adjustbox}
\end{table*}



Table~\ref{tab:results_of_deep_learning_baselines} shows the 
results for all the learning-based approaches.
Compared to \tufano{} and \autotransform, \tooln produces the greatest number of perfect predictions (around 30\%) and more similar code with the perfect predictions (higher value of CodeBLEU). 
Among all baselines, we also observe that the effectiveness of the Vanilla Transformer with localization is similar to \tooln{} with localization (median CodeBLEU=0.911). 

\begin{table}[h]
\centering
\caption{Comparison results with IDD and JDeodorant}
\footnotesize
\label{tab:results_of_other_baselines}
\begin{tabular}{l|rrr} 
\toprule
\multirow{2}{*}{Benchmark} & \multicolumn{3}{c}{Perfect Prediction (\#/\%)}  \\ 
\cline{2-4}
                           & IDD        & JDeodorant & \tooln              \\ 
\hline\hline
Whole(\#9508)              & 1913/20.12 & -          & 2827/29.73            \\ 
\hline
Valid(\#404)               & 82/20.30   & 0/0.00     & 123/30.45             \\
\bottomrule
\end{tabular}
\end{table}

For IDD and \jdeo, we calculate the total number of perfect predictions generated. We did not run \jdeo{} in the whole dataset as it requires compilation. Table~\ref{tab:results_of_other_baselines} shows the results for the two approaches. Overall, \tooln{} outperforms IDD and \jdeo by generating the greatest number of perfect predictions. As \jdeo{} only supported limited types of transformations, our experiment shows that this rule-based approach did not generate any perfect prediction for the \emph{valid} dataset.
Listing~\ref{lst:ppexample}~\cite{ppexample} shows a perfect prediction generated by \tooln{} where other baselines fail to generate the perfect prediction. In this example, \tooln{} applys ``Simplify method return" transformation for simplification, which inlines the variable declaration into the method return, resulting in shorter lines of code.



\begin{lstlisting}[style=mystyle, escapechar=^,caption=A \tooln{} generated perfect prediction example, upquote=true, label={lst:ppexample}]
 public Collection<AuditRequestLog> getAuditRequestLogs() {
-   Collection<AuditRequestLog> newList = repository.findAll();
-   return newList;
+   return repository.findAll();
	}
\end{lstlisting}



\subsection{[RQ6] Simplified Line localization}
\label{sec:localizeresults}
To investigate the effectiveness of the simplified line localization step in \tooln, we evaluate the number of perfect predictions generated by \tooln{} with (\emph{Localized}) and without (\emph{Raw}) simplified line localization. Overall,
we can observe from \emph{Localized} and \emph{Raw} columns in
Table~\ref{tab:results_of_deep_learning_baselines} and Table~\ref{tab:distribution_of_transformation_types} that simplification line localization helps to guide \tooln{} in generating more perfect predictions and increasing the diversity of the transformation types used. Notably, without localization, \tooln{} can only generate 64 perfect predictions but can generate 59 (92\%) more perfect predictions after adding simplified line localization.  
Listing~\ref{lst:localizationexample}~\cite{foreachexample} gives an example showing the differences between program generated with and without line localization. Without simplified line localization, \tooln{} would remove all the \texttt{System.out.println()} in the original method. 
With the line localization step in \tooln{} that marks the first two lines, 
\tooln{} can focus on simplifying these two lines with  ``Use foreach in loop iteration" transformation, leads to the generation of the perfect prediction. 
This example shows that simplified line localization guides \tooln{}
into generating more diverse transformation. 
\begin{lstlisting}[style=mystyle, escapechar=^,caption=Effectiveness of simplified line localization, upquote=true, label={lst:localizationexample}]
- <original>for(int i = 0; i < numbers.size(); i++) {</original>
- <original>Integer currentNum = numbers.get(i);</original>
+ <simplified>for(Integer currentNum : numbers) {</simplified>
    Integer otherNum = map.get(2020 - currentNum);
    if (otherNum != null) {
        System.out.println("this num = " + currentNum);
        System.out.println("other num = " + otherNum);
        int result = otherNum * currentNum;
        System.out.println("result = " + result);
        return result;    } }
\end{lstlisting}






\subsection{[RQ7] Types of Transformations Used}

\begin{table}
\centering
\caption{Diversity of transformation types in perfect predictions in the \emph{valid} dataset}
\label{tab:distribution_of_transformation_types}
\small
\begin{adjustbox}{width=0.48\textwidth}
\begin{tabular}{c|c|c|c|c} 
\toprule
\multirow{2}{*}{Tool} & \multicolumn{2}{c|}{Transformation Type (type (\#case with a specific type))}                                                                                                                                                                                                                                                                                                                                                       & \multicolumn{2}{c}{Total (\#type; \#case)}  \\ 
\cline{2-5}
                      & Raw                                                                                                                                                                                      & Localized                                                                                                                                                                                                       & Raw   & Localized                            \\ 
\hline\hline
Tufano                &          -                                                                                                                                                                          & -                                                                                                                                                                                                               & -   & -                                    \\ 
\hline
AutoTransform         & -                                                                                                                                                                                        & -                                                                                                                                                                                                               & -     & -                                    \\ 
\hline
Vanilla Transformer   & \multicolumn{1}{l|}{T3 (1),~T1.1 (1)}                                                                                                                                                    & \multicolumn{1}{l|}{T3 (1),~T1.1 (1)}                                                                                                                                                                           & 2; 2   & 2; 2                                  \\ 
\hline
\tooln                & \multicolumn{1}{l|}{\begin{tabular}[c]{@{}l@{}}T3 (27),~T1.1 (14), T6.1 (5), \\T2.2 (4), T5.1 (3),~T2.1 (2),\\T7.2 (2), T7.6 (2),~T1.3 (2), \\T1.9 (1),~T1.6 (1),~T7.5 (1)\end{tabular}} & \multicolumn{1}{l|}{\begin{tabular}[c]{@{}l@{}}T3 (72), T1.1 (17), T2.2 (7),~\\T1.10 (6),~T5.1 (6),~T1.3 (4),~\\T2.1 (2),~T6.1 (2),T7.6 (2), \\T7.2 (1), T1.9 (1),~T1.6 (1), \\T7.5 (1),~T4.1 (1)\end{tabular}} & 12; 64 & 14; 123                               \\
\bottomrule
\end{tabular}
\end{adjustbox}
\end{table}

To investigate the diversity of the generated simplifications, we manually analyzed the types of transformation in the generated perfect predictions by the deep learning baselines (we did not evaluate the diversity of $DD^I$ and \jdeo{} as these tools can only support the limited type of transformations). The first two authors of this paper independently checked the transformations; then, they held a meeting to resolve any disagreement. 
Table~\ref{tab:distribution_of_transformation_types} shows the distribution of the transformation types in perfect predictions generated by the deep learning baselines. In the ``Transformation Type'' columns, we use the labels in Table~\ref{tab:rootcause} to refer to the same transformation type discovered in our study, whereas the last two columns represent the total number of transformation types (\#type) used and the total number of perfect predictions generated by each approach for the \emph{valid} dataset. Overall, \tooln{} generates the highest number of perfect predictions using the most diverse set of transformations compared to other baselines. Specifically, it uses 12 (without localization) and 14 (with localization) different types of transformations to generate the simplified programs. We also observe that \tufano{} uses only deletion in both of the correctly generated simplified programs, making this baseline similar to $DD^I$, which performs only deletion correctly. As we observe that all automatically generated programs can be categorized using our taxonomy, this shows that \emph{our taxonomy of simplification is general.} 


\subsection{[RQ8] Quality of the Generated Programs}
To evaluate the quality of the generated programs in the \emph{valid} dataset, we calculate \emph{compilation passing rate and tests passing rate}, which quantifies the ratio of the generated simplified java programs that are test-equivalent. We replace the original method with the simplified method, then compile the whole project and run the test suites. Table~\ref{tab:results_of_compilation_and_tests_execution} shows the compilation success ratio and test-equivalent ratio on our valid dataset for each deep-learning-based baseline. Overall, the programs generated by \tooln are of higher quality with respect to compilation passing rate and test passing rate. Specifically, \tooln performs better with line localization, which is almost double the ratio with respect to compilation passing rate (35.40\%) and tests passing rate (31.19\%). Compared to other baselines, \tooln could generate simplified programs that are more likely to have the right syntax and keep the original semantics.

We also measure the quality of the simplified programs generated by \tooln from three other metrics: SLOC (Source Lines of Code), Cyclomatic Complexity, and Readability. We measure the perfect predictions and semantic-equivalent simplifications generated by \tooln. As other baselines only generate very few numbers of perfect predictions and semantic-equivalent (e.g., Vanilla Transformer can only generate 4 semantic-equivalent program and zero perfect prediction), we compare our tool with the ideal delta debugging (IDD) only, which generates 82 (20.30\%) perfect predictions. Table \ref{tab:results_of_other_metrics} provides the results with respect to these three metrics. Specifically, for SLOC, \tooln performs better on the raw data, and the generated semantic-equivalent simplifications achieve the highest reduction ratio (18.13\%) on the average number of SLOC.

\begin{table}
\centering
\caption{Quality of generated programs}
\label{tab:results_of_compilation_and_tests_execution}
\small
\begin{adjustbox}{width=0.45\textwidth}
\begin{tabular}{l|rr|rr} 
\toprule
\multirow{2}{*}{Tool} & \multicolumn{2}{c|}{\begin{tabular}[c]{@{}c@{}}Compilation Success\\\#/\%\end{tabular}} & \multicolumn{2}{c}{\begin{tabular}[c]{@{}c@{}}Tests-equivalent\\\#/\%\end{tabular}}  \\ 
\cline{2-5}
                      & Raw               & Localized                                                           & Raw               & Localized                                                        \\ 
\hline\hline
Tufano                & 22/5.45           & 10/6.19                                                             & 16/3.96            & 7/1.73                                                           \\ 
\hline
AutoTransform         & 2/0.50            & 2/0.50                                                              & 2/0.50            & 2/0.50                                                           \\ 
\hline
Vanilla Transformer   & 4/0.99            & 4/0.99                                                              & 4/0.99            & 4/0.99                                                           \\ 
\hline
\tooln                & \textbf{70/17.33} & \textbf{143/35.40}                                                  & \textbf{63/15.59} & \textbf{126/31.19}                                               \\
\bottomrule
\end{tabular}
\end{adjustbox}
\end{table}

\begin{table*}
\centering
\caption{Simplification results with respect to SLOC, Complexity, and Readability}
\label{tab:results_of_other_metrics}

\begin{adjustbox}{width=0.8\linewidth,center}
\begin{tabular}{c|c|cc|cc|cc|cc|cc} 
\toprule
\multirow{4}{*}{Metric}      & \multirow{4}{*}{Type} & \multicolumn{2}{c|}{\multirow{2}{*}{IDD}} & \multicolumn{8}{c}{\tooln}                                                                                                       \\ 
\cline{5-12}
                             &                       & \multicolumn{2}{c|}{}                     & \multicolumn{4}{c|}{~Raw}                                 & \multicolumn{4}{c}{Localized}                                        \\ 
\cline{3-12}
                             &                       & \multicolumn{2}{c|}{PP}                   & \multicolumn{2}{c|}{PP}         & \multicolumn{2}{c|}{SE} & \multicolumn{2}{c|}{PP}          & \multicolumn{2}{c}{SE}            \\ 
\cline{3-12}
                             &                       & O     & S                                 & O    & S                        & O     & S               & O     & S                        & O     & S                         \\ 
\hline
\multirow{3}{*}{SLOC}        & mean                  & 14.66 & 12.66 (-13.64\%)                  & 8.77 & 7.18 (-\textbf{18.13}\%) & 10.65 & 9.13 (-14.27\%) & 12.12 & 10.19 (-15.92\%)         & 12.39 & 10.23 (-17.43\%)          \\
                             & median                & 13.00 & 10.00 (-23.08\%)                  & 8.00 & 6.50 (-18.75\%)          & 10.00 & 8.00 (-20.00\%) & 11.00 & 8.00 (-\textbf{27.27}\%) & 11.00 & 9.00 (-18.18\%)           \\
                             & st. dev.              & 7.88  & 7.45                              & 4.39 & 4.18                     & 5.99  & 5.69            & 6.57  & 6.36                     & 6.76  & 6.46                      \\ 
\hline
\multirow{3}{*}{Complexity}  & mean                  & 3.06  & 2.83~(-7.52\%)                    & 1.95 & 1.91 (-2.05\%)           & 2.56  & 2.35~(-8.20\%)  & 2.63  & 2.35~(-10.65\%)          & 3.13  & 2.63~(-\textbf{15.97}\%)  \\
                             & median                & 2.0   & 2.0~(-0.00\%)                     & 1.00 & 1.00 (-0.00\%)           & 2.00  & 2.00~(-0.00\%)  & 2.00  & 2.00 (-0.00\%)           & 2.50  & 2.00 (-\textbf{20.00}\%)  \\
                             & st. dev.              & 2.61  & 2.44                              & 1.43 & 1.44                     & 1.70  & 1.52            & 2.07  & 1.83                     & 2.36  & 2.02                      \\ 
\hline
\multirow{3}{*}{Readability} & mean                  & 2.48  & 2.26~(-8.87\%)                    & 1.09 & 1.05~(-3.67\%)           & 2.11  & 1.87 (-11.37\%) & 2.04  & 1.76~(-13.73\%)          & 2.90  & 2.37~(-\textbf{18.28}\%)  \\
                             & median                & 1.0   & 1.0 (-0.00\%)                     & 0.50 & 0.00~(-\textbf{100}\%)   & 1.00  & 1.00 (-0.00\%)  & 1.00  & 1.00 (-0.00\%)           & 2.00  & 1.00 (-\textbf{50.00}\%)  \\
                             & st. dev.              & 3.36  & 3.27                              & 1.34 & 1.36                     & 2.60  & 2.26            & 2.90  & 2.77                     & 3.68  & 3.33                      \\
\bottomrule
\end{tabular}
\end{adjustbox}
\begin{tablenotes}
\footnotesize
\item{
SLOC: source lines of code; PP: perfect prediction; SE: semantic-equivalent; O: original program; S: simplified program.
}
\end{tablenotes}
\end{table*}

\section{Threats to Validity}
We identify the following threats to validity of this paper:

\noindent \textbf{External.} The set of program transformations used by developers found in our study merely represents transformations that are frequently occurred in these commits (i.e., we do not claim that our identified set is a ``complete'' set). We mitigate this by studying \simpli{} from two perspectives: (1) PRs in the study, and (2) commits in our mined dataset. 
For verifiability, we also release the generated programs. 
As our study and evaluation only focus on Java open-source projects, the findings may not generalize beyond Java and other closed-source projects. 

\noindent\textbf{Internal.} Our scripts and tool may have bugs that can affect our results. To mitigate this, we share our results, dataset and tool. 

\noindent\textbf{Conclusion.} Conclusion threats include (1) overfitting of our dataset and (2) subjectivity of manual analysis. We minimize (1) by ensuring that the training and testing dataset use data from different projects. We mitigate (2) by cross-validating between two annotators.  

\section{Implication}

Based on our study and evaluation, we discuss the implications for developers and researchers.

\noindent \textbf{Implication for Developers.} Our study identifies a set of commonly used program transformations for \simpli{} and developers' motivations. Based on our study's two most common transformations (Extract method and \useapi), we realized that developers tend to perform simplification by introducing new methods or reusing existing API (Finding 1). 
Meanwhile, despite sharing the same goals of obtaining a simplified program, some IDE-supported simplifications need to be invoked under different menus (Finding 1). As IDE users have expressed difficulty in finding these transformations~\cite{organize,diamond}, IDE plugin developers should consider \emph{enhancing the usability of the by aggregating all transformations} that share the same purpose of simplification. For example, most modern IDE supports ``Remove unused imports'' as an ``Organize imports'' feature, whereas ``Use diamond operator'' is available only under ``Java inspection'' in IntelliJ. Our study of the motivations behind \simpli{} suggests developers to \emph{use existing static metrics to ensure the quality of the manually crafted simplified programs} (Finding 2).
Moreover, although most simplification types are covered by existing refactoring types (Finding 3), prior auto-refactoring tools only support limited types (Finding 4). This means that developers can only use prior auto-refactoring tools (e.g., \jdeo{}) for the few supported types. Hence, we propose \tooln to automate \simpli{}. 

\noindent \textbf{Implication for Researchers.}
Our study and proposed program simplification framework lay the foundation for research in three promising directions. 
First, our study \emph{provides the key criteria that drive the design of future automated tools for \simpli{}}. Such tools should (1) contain a richer set of program transformations (Finding 1), (2) incorporate language-specific features (Finding 1), (3) incorporate prior static metrics (e.g., SLOCs, readability and cyclomatic complexity) for checking the quality of the generated code (Finding 2). Our proposed framework and the \bench{} dataset lay the first step in promoting future research towards this direction. 
Second, \emph{there exists limitations of adapting existing refactoring detection and auto-refactoring tools for detecting and automating \simpli{}}. As \simpli{} involves diverse types of program transformations, our study reveals that several frequently used transformations, e.g., replacing with equivalent API and simplifying expression, have not been automated (Finding 1). To enhance the usability of IDE-supported simplifications, it is worthwhile to investigate better UI design to help users to invoke different transformations that share the same goals of simplifications (Finding 3). Within the supported refactoring types, refactoring detection and auto-refactoring tools fail to detect them due to the limitations in the hand-crafted rules used for detection (Finding 4) and limited types of refactoring supported by auto-refactoring tools (Finding 5). 
Third, \emph{our taxonomy of \simpli{} serves as preliminary studies for motivating future research on using a richer set of transformations for improving deletion-based approaches} (e.g., program reduction and debloating). As deletion-based approaches have been applied in many domains (Section~\ref{sec:related}), it is worthwhile future work to investigate how to incorporate more human-like transformations to further improve these deletion-based techniques. For example, instead of relying on genetic operators (mutation and crossover), incorporating our simplifications to remove code bloat in genetic
programming
 could help generate more diverse programs.   
\section{Conclusion}
We present the first study of \simpli{} in OSS projects, focusing on the transformations types, developers' motivations, and the transformations covered by existing refactoring types. Our study reveals gaps in applying existing approaches for automating \simpli{} and outlines the criteria for designing automatic simplification techniques. 
Based on our study, we propose \tooln{}, an automated simplification framework that learns from our dataset \bench{} using simplified line localization, and a checker. Our experiment shows that \tooln{} is more effective than prior approaches. 

\section*{Data Availability}
The data and code are available at ~\cite{dataandcodelink}.

\bibliography{refs}

\end{document}